\documentclass[useAMS,usenatbib]{mn2e}

\usepackage{epsfig,  bm}

\newcommand{\beq}{\begin{equation}}
\newcommand{\eeq}{\end{equation}}
\newcommand{\bea}{\begin{eqnarray}}
\newcommand{\eea}{\end{eqnarray}}
\newcommand{\trm}[1]{\textrm{#1}}

\newcommand       \apj          {ApJ}

\newcommand       \aap          {A\&A}

\newcommand       \mnras        {MNRAS}

\title[Saturation of g-modes in proto-neutron stars]{Non-linear saturation of \lowercase{g}-modes in proto-neutron stars: quieting the acoustic engine}

\author[N. N.~Weinberg and E.~Quataert]{Nevin N.~Weinberg,$^{1}$\thanks{E-mail:nweinberg@astro.berkeley.edu} Eliot Quataert$^{1}$\\
$^{1}$Astronomy Department and Theoretical Astrophysics Center,
University of California, Berkeley, 601 Campbell Hall, Berkeley CA,
94720}

\begin{document}

\date{Accepted. Received; in original form}

\pagerange{\pageref{firstpage}--\pageref{lastpage}} \pubyear{2008}
\maketitle

\label{firstpage}

\begin{abstract}

According to Burrows et al.'s acoustic mechanism for core-collapse supernova explosions, the primary, $l=1$, g-mode in the core of the proto-neutron star is excited to an energy of $\sim 10^{50}\trm{ ergs}$ and damps by the emission of sound waves. Here we calculate the damping of the primary mode by the parametric instability, i.e., by nonlinear, 3-mode coupling between the low-order primary mode and pairs of high-order g-modes. We show that the primary mode is strongly coupled to highly resonant, neutrino damped pairs with $n\ga 10$; such short wavelength interactions cannot be resolved in the simulations. We find that the parametric instability saturates the primary mode energy at $\sim 10^{48}\trm{ ergs}$, well below the energy needed to drive an explosion. We therefore conclude that acoustic power is unlikely to be energetically significant in core-collapse supernova explosions. 
\end{abstract}

\begin{keywords}
\vspace{-0.4cm}
hydrodynamics --- instabilities; stars: neutron --- supernovae: general
\end{keywords}

\section{Introduction}
\label{sec:intro} 

Of the $\sim 10^{53}\trm{ ergs}$ of gravitational binding energy released in a stellar core-collapse, $\sim 1\%$ is used to power the supernova explosion and the remainder is nearly all lost to neutrinos that escape to infinity. According to the neutrino mechanism, the supernova is driven by the small fraction of neutrinos that deposit energy in the gain region just below the stalled accretion shock. Over 20 years of study has demonstrated that the viability of the neutrino mechanism hinges on the multidimensional nature of core-collapse (see e.g., \citealt{Janka:07} for a review), with some recent 2D simulations on the borderline of success \citep{Buras:06, Marek:07}. 
 
Recently, Burrows et al. (2006, 2007) proposed a new explosion mechanism that instead relies on acoustic power generated in the core of the proto-neutron star (PNS).  Using 2D, axisymmetric, multigroup, radiation-hydrodynamic simulations, Burrows et al. find that after the stalled shock becomes unstable to the standing accretion shock instability (SASI; see \citealt{Blondin:06, Foglizzo:07}), anisotropic accretion streams excite and maintain vigorous g-mode oscillations in the PNS core. Sound pulses radiated from the core steepen into shock waves and deposit sufficient energy and momentum in the outer mantle to drive an asymmetric explosion $\approx 1 \trm{ s}$ after bounce. 

The primary mode of oscillation seen by Burrows et al. is a low-order core g-mode with spherical degree $l=1$ (frequency $\approx 300 \trm{ Hz}$). For an $11 M_\odot$ progenitor, they find that the pulsational energy in the core is maintained at $\sim 10^{50}\trm{ ergs}$ for several hundred milliseconds before the explosion and that the dominant source of damping is the emission of sound waves near the PNS surface. For more massive progenitor models, the core pulsations achieve energies near $\sim 10^{51}\trm{ ergs}$. 

As the primary mode is driven to ever larger amplitude, nonlinear coupling to other PNS modes can become an important source of damping. Here we calculate the leading-order nonlinear coupling between the primary mode and high-order core g-modes (the parametric instability). The latter rapidly damp due to neutrino diffusion. Burrows et al.'s resolution in the PNS core is $\approx 0.5 \trm{ km}$ and they cannot adequately resolve the coupling of the primary to higher order modes; an analytic calculation is thus required. 

We begin in \S~\ref{sec:saturation} with a brief review of the parametric instability and outline the calculation of the saturation energy. We then describe our PNS model and its eigenmodes (\S~\ref{sec:proto}), their coupling strength (\S~\ref{sec:coupling}), and their driving and damping rates (\S\S~\ref{sec:drivingrate}, \ref{sec:dampingrates}). In  \S~\ref{sec:results} we show results and discuss their implications.  A more detailed presentation of our methods and results will be presented in a separate paper.

\section{Saturation by nonlinear mode coupling}
\label{sec:saturation} 

In the particular context of three-mode coupling considered here, if the amplitude of the $l=1$ ``parent" mode exceeds the parametric instability threshold, it begins to transfer energy to two lower-frequency ``daughter" modes (\citealt{Dziembowski:82, Kumar:96}; \citealt{Wu:01}, hereafter WG01; \citealt{Arras:03}). The daughter mode amplitudes then rise exponentially, and the system has an equilibrium in which the parent mode energy is
\beq
\label{eq:E1}
E_{1, {\rm eq}} = \frac{\gamma_2 \gamma_3}{18 \kappa^2 \omega_2 \omega_3}\left[1 + \left(\frac{2 \delta \omega}{\gamma_2 + \gamma_3-\gamma_1}\right)^2\right].
\eeq
Here $\omega_i$ is the eigenfrequency of mode $i$ (an index of 1 refers to the parent mode and 2 and 3 to the two daughter modes), $\delta \omega=\omega_1 - \omega_2 - \omega_3$ is the detuning frequency, $\gamma_{2,3}$ are the daughter damping rates, $\gamma_1$ is the parent driving rate, and $\kappa$ is the nonlinear coupling coefficient.\footnote{We follow the normalization conventions of WG01.}  The daughter modes' equilibrium energy is $E_{2,3, {\rm eq}}/E_{1, {\rm eq}} = Q_{2,3} \gamma_1/\omega_1$, where $Q_{2,3}=\omega_{2,3}/\gamma_{2,3}$ is the daughter mode quality factor. As we describe in \S~\ref{sec:results}, the parent's saturation energy $E_{\rm sat}$ is set by the daughter pairs that minimize $E_{1, {\rm eq}}$. 

Accretion onto the PNS pumps energy into the primary mode at a rate $\gamma_1=\dot{E}_1/E_1$, where according to Burrows et al. $\dot{E}_1 \sim10^{50}-10^{51}\trm{ ergs s}^{-1}$. Unlike in many other applications of the parametric instability (e.g., \citealt{Arras:03}), the driving here is nonlinear because $\dot E_1$, rather than $\gamma_1$, is taken to be constant. Equation (\ref{eq:E1}) is thus a cubic equation in $E_1$. For $|\delta \omega|/\omega_1 \la 0.01$, it has three real roots if $\dot{E}_1 / (\gamma_2 +\gamma_3) \ga [18 \kappa^2 Q_2 Q_3]^{-1}$ and one real root otherwise. In the three root case, the largest root $E_{1, {\rm eq}} \approx \dot{E}_1 / (\gamma_2 +\gamma_3) $ is the stable solution (as revealed by both an analytic stability calculation and direct numerical solution of the time-dependent amplitude equations for the 3-mode system). In the one root case, the solution for small detuning is $E_{1, {\rm eq}}\approx [18 \kappa^2 Q_2 Q_3]^{-1}$. 

\begin{figure}
\begin{center}
\epsfig{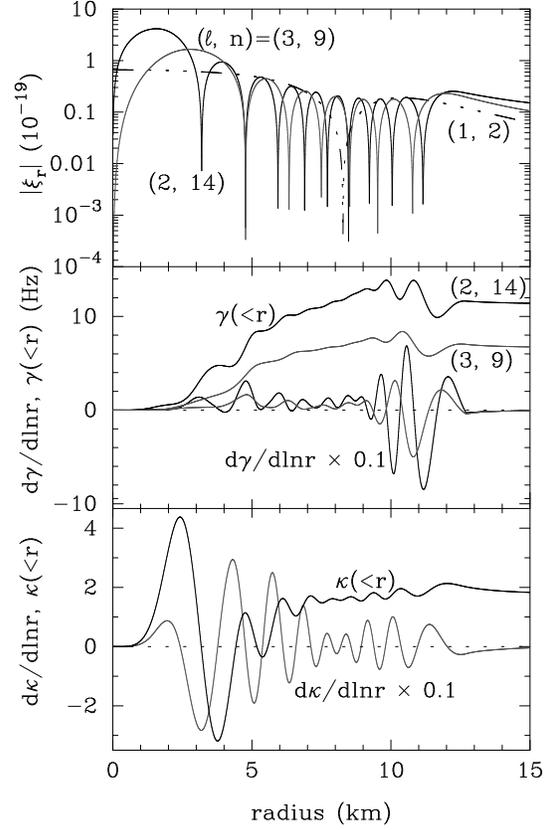}
\end{center}
\caption{Profiles of the radial displacement $\xi_r$ (\emph{top panel}; normalized such that the total pulsational energy is one), daughter damping rates (\emph{middle panel}; $\gamma(<r)=\int_0^r (d\gamma/dr)\, dr$), and coupling coefficient  (\emph{bottom panel}; $\kappa(<r)=\int_0^r (d\kappa/dr)\, dr$, in units of $10^{-26}\trm{ ergs}^{-1/2}$) for 3-mode coupling of the primary mode to a pair of modes with $(l,n)=(2, 14)$ and $(3, 9)$. Table \ref{tab:tab1} lists additional properties of this triplet. \label{fig:singlemode}}
\end{figure}

Before describing the calculation of $\kappa$ and $\gamma$, we first give an estimate of $E_{\rm sat}$. In our units, the total energy of a mode is normalized to unity and the coupling coefficient for core g-modes is of order $\kappa \sim 1/\sqrt{E_{\rm bind}}=(GM^2/R)^{-1/2}\approx (10^{53} \trm{ ergs})^{-1/2}$, the inverse square-root of the binding energy of a PNS of mass $M$ and radius $R$. The strength of the 3-mode coupling is sensitive to the structure of the modes, particularly the turning points and the location of nodes. While for most daughter pairs the coupling is weak ($\kappa \ll 1/\sqrt{E_{\rm bind}}$), we show that some pairs couple strongly to the primary, with $\kappa \approx 10/ \sqrt{E_{\rm bind}}$. Of the strongly coupled pairs a few with $n/l\sim5$ have damping rates $\gamma_{2, 3}\approx 100\trm{ Hz}$ and $|\delta \omega|/\omega_1 \sim 0.01$. For these fiducial daughter modes, if $\dot{E}_1\la10^{50}\trm{ ergs s}^{-1}$, the primary saturates at an energy of
\beq
\label{eq:Esat}
E_{\rm sat} \sim 10^{48}
\left(\frac{\kappa}{10\sqrt{R_{30}/M^2_{1.4}}}\right)^{-2}
\left(\frac{Q_{2,3}}{10}\right)^{-2}\trm{ ergs} ,
\eeq
($R_{30}=R/30 \trm{ km}$ and $M_{1.4} = M/1.4 M_\odot$) while for larger $\dot{E}_1$ the three root solution to equation (\ref{eq:E1}) applies and $E_{\rm sat}\sim5\times10^{48}(\dot{E}_1/10^{51}\trm{ ergs s}^{-1})\trm{ ergs}$. 

After the parent crosses the parametric threshold, the energy growth rate of the daughter modes is $\sigma_{2,3} \approx [18E_1 \kappa^2 \omega_2 \omega_3-(\delta\omega)^2]^{1/2}-\gamma_{2,3}$. Thus, if the parent overshoots its equilibrium by just a factor of two ($E_1\simeq2E_{1, {\rm eq}}$), $\sigma\simeq100\trm{ Hz}$ and the daughters undergo many tens of e-foldings of growth in just a small fraction of a second. Although the initial daughter mode amplitudes are uncertain, we expect PNS convection (and perhaps also the asymmetric accretion streams) to excite them to energies $\gg 10^{40} \trm{ ergs}$. Equilibrium [equations (\ref{eq:E1}) \& (\ref{eq:Esat})] is therefore established in $\la 0.1\trm{ sec}$. At the energies found in Burrows et al.'s simulations ($E_1\sim10^{50}\trm{ ergs}$), the daughters would grow in $\sim 1 \trm{ ms}$ and very quickly drain energy out of the primary. 

\subsection{Proto-Neutron Star Model and Eigenmodes}
\label{sec:proto}

The PNS models used in our study were provided by A. Burrows. They are angle-averaged radial slices of density ($\rho$), temperature ($T$), electron fraction ($Y_e)$, and pressure ($p$), at various times during their core-collapse simulation of the $15M_\odot$ \citet{Woosley:02} progenitor model (see \citealt{Burrows:07}). We show results for the PNS at $t=500 \trm{ ms}$ after core bounce, although the value of $E_{\rm sat}$ is nearly constant in time.\footnote{We also obtain $E_{\rm sat}\approx 10^{48} \trm{ ergs}$ for a PNS model taken from a 1D core collapse simulation of an $11 M_\odot$ progenitor (provided by T. Thompson; see \citealt{Thompson:05}).}

We interpolate over the model and compute the adiabatic oscillation frequencies and eigenfunctions using the non-relativistic Aarhus adiabatic oscillation package \citep{Christensen:07}. We use the \citet{Lattimer:91} equation of state (EOS) to calculate the Brunt-V\"ais\"al\"a (buoyancy) frequency $N$, chemical potentials $\mu$, etc., needed in the eigenfunction and damping calculations. Since the modes we consider all have damping rates $\gamma \ll \omega$, the adiabatic approximation is appropriate. The eigenfunctions and coupling coefficient are computed in the Cowling approximation (i.e., we ignore perturbations to the gravitational potential), which is appropriate for the high-order daughter modes under consideration and should also be reasonable for the parent mode. 
We assume that the Lagrangian pressure perturbation vanishes at the outer boundary, which we place at $\rho=10^9\trm{ g cm}^{-3}$ (corresponding to a radius  $r_{\rm outer} \simeq 80\trm{ km}$).  Our results are not sensitive to the location of the outer boundary since the nonlinear coupling is dominated by the core ($r < 15 \trm{ km}$) and the mode damping is dominated by the core and the region near the neutrinosphere ($r\approx 40\trm{ km}$).

\begin{table}
\caption{Daughter pairs with low parametric thresholds}
\label{tab:tab1}
\begin{tabular}{ccccc}
\hline
$(l_2, n_2):(l_3, n_3)$ &$|\delta \omega| / \omega_1$ & $\gamma_2, \gamma_3$ & $|\kappa|$ & $E_{1, {
\rm eq}}$(ergs)$^{\rm a}$\\
\hline
$(8, 39):(9, 33)$  & $0.5$ & $132, 69$ & $1.2$ & $5.1\times10^{48}$\\
$(3, 21):(4, 13)$  & $2.5$ & $740, 25$ & $1.3$ & $7.2\times10^{48}$\\
$(1, 16):(2, 4)$  & $16.2$ & $548,12$ & $0.9$ & $8.0\times10^{48}$\\
$(2, 14):(3, 9)^{\rm b}$  & $5.7$ & $241, 250$ & $1.8$& $1.3\times10^{49}$\\
\hline
\end{tabular}

Note.---$|\delta \omega| / \omega_1$ in $10^{-3}$; $\gamma_2, \gamma_3$ in Hz; $|\kappa|$ in  $10^{-
26} \trm{ ergs}^{-1/2}$.\\
$^{\rm a}$ Values of $E_{1, {\rm eq}}$ are for $\dot{E}_1=10^{51}\trm{ ergs s}^{-1}$.\\
$^{\rm b}$ See Fig. \ref{fig:singlemode} for $\xi_r(r)$, $\gamma(r)$, and $\kappa(r)$ for this triplet.
\\ 
\vspace{-0.1cm}
\end{table}

Motivated by Burrows et al.'s results, we assume that the primary (i.e., parent) mode has a spherical degree $l_1=1$ and radial order $n_1=2$. We find that it is a core mode with frequency $\omega_1=2\pi\times297\trm{ Hz}$, in good agreement with Burrows et al..  Although we use the Lattimer-Swesty EOS instead of the \citet{Shen:98} EOS used by Burrows et al., this agreement is reasonable given that we use Burrows et al.'s $\{\rho, T , Y_e , p\}$ profiles; i.e., the Brunt-V\"ais\"al\"a frequency of the two EOS's---which largely determines $\omega$---differ by $\delta \Gamma_1 / 2\Gamma_1 \la 10\%$, half the fractional difference of the EOSs' adiabatic index $\Gamma_1$ (see \citealt{Rosswog:02}). Although the $l=1$, $n=1$ mode has a similar frequency ($324 \trm{ Hz}$), it is a surface mode that lies almost entirely above the PNS convection zone ($r\ga 20\trm{ km}$; similar to Fig. 4 of \citealt{Yoshida:07}).  We find that it has a neutrino damping rate of $\sim 10^3 \trm{ Hz}$. It therefore seems inconsistent with Burrows et al.'s primary mode and we instead focus on the $n=2$ mode.\footnote{We find that the $n=1$ mode also has $E_{\rm sat}\sim 10^{48}\trm{ ergs}$.}

We consider the 3-mode coupling of the primary to all daughter modes with $1\le l \le 10$ and frequencies in the range $\omega/\omega_1\approx[1/4, 3/4]$ (corresponding to $3\le n \le18$ for $l=1$ and $30\le n \le64$ for $l=10$). Selection rules enforce conservation of angular momentum ($|l_2-l_3|\le l_1 \le l_2+l_3$ with $l_1+l_2+l_3$ even; see WG01), leaving $\simeq 6000$ triplets with nonvanishing $\kappa$. The top panel of Figure \ref{fig:singlemode} shows the radial displacement $\xi_r(r)$ of the primary mode and a strongly-coupled, resonant daughter pair. 

\subsection{Coupling Coefficients}
\label{sec:coupling}

The three-mode coupling coefficient for adiabatic modes under the Cowling approximation is given by (\citealt{Kumar:89, Kumar:96}; WG01; \citealt{Schenk:02})
\bea
\label{eq:kappa}
\kappa &=& -\frac{1}{6}\int d^3x \, p \Biggl\{\left((\Gamma_1 -1)^2 + \left. \frac{\partial \Gamma_1}{\partial \ln\rho}\right|_{\rm ad} \right)\left(\bm{\nabla}\bm{\cdot}\bm{\xi}\right)^3
\nonumber\\ & &
+3(\Gamma_1-1)\left(\bm{\nabla}\bm{\cdot}\bm{\xi}\right)\xi^i_{;j}\xi^j_{;i}
+2\xi^i_{;j}\xi^j_{;k}\xi^k_{;i}\Biggr\}
\nonumber\\ & &
+\frac{1}{6}\int d^3x \,\rho \xi^i \xi^j \xi^k \phi_{;ijk}\,,
\eea
where $p$, $\rho$, and $\phi$ are the unperturbed pressure, density, and gravitational potential, $\Gamma_1$ is the adiabatic index, and $\bm{\xi}$ is the Lagrangian displacement vector. The semicolon denotes covariant derivative and the integration is over the volume of the star. As in appendix A1 of WG01, we carry out the angular integrations analytically and use integration by parts to put this expression in a form more suitable for numerical computation.\footnote{There are errors in terms 5, 7, and 9 of WG01's final expression (their eq. [A15]; the errors are also present in eq. [A14]). We correct these and also include the unperturbed gravity term (last term in our eq. [\ref{eq:kappa}]), which they neglect.} The bottom panel of Figure \ref{fig:singlemode} shows $d\kappa/d\ln r$ and $\kappa(<r)=\int_0^r (d\kappa/dr)\, dr$ for the coupling of the primary mode to a strongly coupled, resonant daughter pair.

The strength of the coupling depends on the triple product of the covariant derivative of each mode's eigenfunction and is very sensitive to the location of nodes and turning points (i.e., where $N=\omega$). We find that even after we carry out the angular integrations analytically and use integration by parts, the modes need to be resolved to $\la 0.1\trm{ km}$ in order to accurately calculate $\kappa$. If we instead calculate $\kappa$ by directly integrating equation (\ref{eq:kappa}) over a cartesian grid, a resolution  of $\la 0.01\trm{ km}$ is needed. It is difficult to infer from these analytic techniques what resolution a hydrodynamic simulation must have in order to accurately capture the nonlinear interaction; we nonetheless believe that the resolution of current simulations ($\ga 0.5\trm{ km}$) is not sufficient. 

If daughter modes parametrically excite granddaughter modes before reaching equilibrium with the primary, they may saturate at energies below $E_{2, 3, {\rm eq}}$ (see e.g., \citealt{Kumar:96}; WG01). By searching the $(l, n)$ parameter space for granddaughter modes that can couple to daughter modes, we find that daughters with $l<4$ do not excite granddaughter modes at their equilibrium energy $E_{2, 3, {\rm eq}}$; for higher $l$ daughter modes there generally are pairs of granddaughters that can be parametrically excited. Accounting for such coupling requires a full mode network calculation which we leave to a future paper. As can be inferred from Table \ref{tab:tab1}, however, when we limit our calculation to daughters with $l<4$, we still find that $E_{\rm sat}\sim 5\times10^{48}\trm{ ergs}$. 

\subsection{Primary Driving}
\label{sec:drivingrate}

Burrows et al. find that accretion pumps energy into the primary mode at a rate $\dot{E}_1\sim10^{50}-10^{51}\trm{ ergs s}^{-1}$. In solving the time-dependent coupling between the parent and daughters, we thus assume that the parent is driven with a constant energy injection rate of $\dot{E}_1 = 10^{50}\trm{ ergs s}^{-1}$ or $\dot{E}_1 = 10^{51}\trm{ ergs s}^{-1}$.  A constant $\dot{E}_1$ corresponds to a nonlinear driving rate $\gamma_1 = \dot{E}_1/E_1$ in equation (\ref{eq:E1}).

Since the accretion streams that excite the primary are generated by the nonlinear SASI oscillation, an alternative modeling approach is to assume a constant linear driving rate $\gamma_1$ equal to the SASI oscillation frequency. According to \citet{Burrows:07}, the frequency of the dominant, time-averaged, nonlinear SASI oscillation ranges from $30 \trm{ Hz}$ (for the $11.2M_\odot$ progenitor model) to $80\trm{ Hz}$ (the $25 M_\odot$ model). As can be inferred from equation (\ref{eq:E1}), such a linear driving model (e.g., with $\gamma_1= 50\trm{ Hz}$) yields very similar results to the nonlinear driving model. 

\subsection{Damping Rates}
\label{sec:dampingrates}
In the neutrino-degenerate PNS core, the energy damping rate of the daughter modes, $\gamma_{2, 3}$, is dominated by neutrino damping due to conduction and particle diffusion (see e.g., \citealt{vandenHorn:84}).\footnote{Turbulent viscosity in the convection zone increases the damping rates by at most $\approx30\%$ and we therefore ignore it here.}  The modes are evanescent in the PNS convection zone ($13\trm{ km} \la r \la 20 \trm{ km}$) and daughter modes with wavelengths $\lambda$ much shorter than the size of the convection zone have negligible amplitudes outside of the core. By contrast, longer wavelength daughter modes tunnel through the convection zone and have a non-negligible amplitude near the neutrinosphere at $r_\nu \approx40\trm{ km}$. Such modes experience significant additional damping due to optically thin emission in the region $r_{\rm thin} \la r \la r_\nu$ where the neutrino mean free path $\langle d_\nu \rangle > \lambda$ ($r_{\rm thin}$ is the radius where $\langle d_\nu \rangle=\lambda$). Modes with $\lambda \sim r_\nu - r_{\rm thin} \approx few\trm{ km}$ experience significant optically thin damping at these radii, while longer wavelength modes do not.

We account for damping in the core and near the neutrinosphere as follows. We assume that the damping occurs in the quasi-adiabatic limit such that $\gamma =-\langle\dot{E}\rangle / E_{\rm tot}$ where $\langle\dot{E}\rangle$ is the time-averaged work integral expression for the energy lost and $E_{\rm tot}$ is the total pulsational energy. We calculate the Rosseland mean free path of the neutrinos $\langle d_\nu \rangle$ using cross-sections given in \citet{Reddy:98}. We assume $\gamma=\gamma_{\rm thick}+\gamma_{\rm thin}$, where
\beq
\label{eq:gthick}
\gamma_{\rm thick} = 
\frac{1}{2E_{\rm tot}}\int_0^{r_{\rm thin}} d^3x \, \rho \frac{\delta T}{T} \delta \!\left( - \frac{1}{\rho}\bm{\nabla}\bm{\cdot}\bm{F}_\nu + \frac{\mu_\nu}{\rho}\bm{\nabla}\bm{\cdot}\bm{N}_\nu\right)
\eeq
accounts for optically thick damping in the core and 
\beq
\label{eq:gthin}
\gamma_{\rm thin} = 
 \frac{1}{2E_{\rm tot}}\int_{r_{\rm thin}}^{r_{\rm outer}} d^3x \, \rho \frac{\delta T}{T} \delta \epsilon
\eeq
accounts for optically thin damping near the neutrinosphere. In equations (\ref{eq:gthick}) \& (\ref{eq:gthin}), $\delta$ denotes a Lagrangian perturbation, $\bm{N}_\nu = -a_T \bm{\nabla} T - a_\eta \bm{\nabla}\eta$ and $\bm{F}_\nu = -b_T \bm{\nabla} T - b_\eta \bm{\nabla} \eta$ are the neutrino number flux and energy flux with transport coefficients $\{a_T, a_\eta, b_T, b_\eta\}$ (see e.g., \citealt{Miralles:00}), $\eta = \mu_\nu / kT$ is the neutrino degeneracy parameter, and $\mu_\nu$ is the neutrino chemical potential. The neutrino emissivity $\epsilon$ is due to charge-current processes and the production of electron, muon, and tau neutrino pairs,  which we calculate using the formulae in Thompson et al. [2001; equations (20) \& (29)] and Thompson et al. [2000; equation (53)]. 

The middle panel of Figure \ref{fig:singlemode} shows $d\gamma/d\ln r$ and $\gamma(<r)=\int_0^r (d\gamma/dr)\, dr$ in the core for a strongly-coupled, intermediate-order, resonant daughter pair. The majority ($\approx 95\%$) of the damping for this pair occurs outside the core, in the optically thin region (see Table \ref{tab:tab1}). By contrast, we find that the damping rate of the highest-order modes ($n\ga25$) is dominated by the core and is to a good approximation, $\gamma \approx \gamma_{\rm thick} \approx 0.1 \left(n^2 + l^2\right)\trm{ Hz}$; this is consistent with a PNS with a $\approx 10\trm{ s}$ Kelvin-Helmholtz cooling time. 

\begin{figure}
\begin{center}
\epsfig{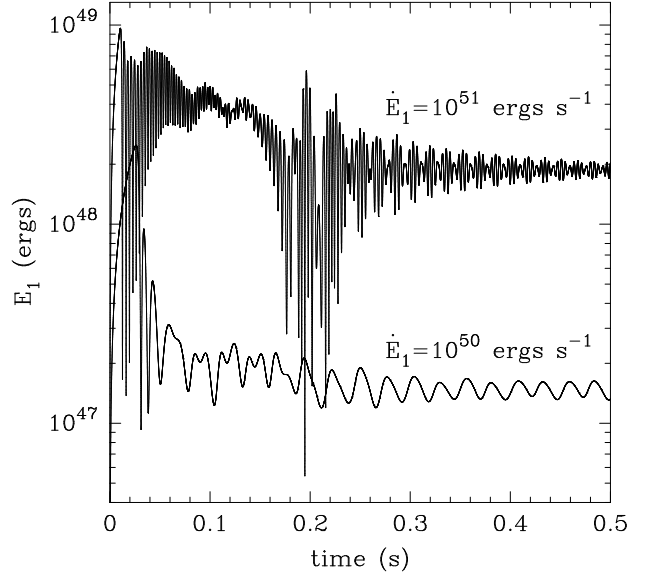}
\end{center}
\caption{Energy of the driven primary $(l=1, n=2)$ g-mode in a PNS as a function of time for coupling between the primary and all g-modes with $1\le l \le 10$ and $1/4 \la \omega/\omega_1 \la 3/4$  (not including daughter only or granddaughter couplings). The primary mode is assumed to be driven with a constant energy injection rate of $\dot{E}_1=10^{50}\trm{ ergs s}^{-1}$ (\emph{bottom curve}) and  $\dot{E}_1=10^{51}\trm{ ergs s}^{-1}$ (\emph{top curve}), corresponding to the power supplied by accretion seen in the simulations of Burrows et al. The initial energy of the daughters is $10^{43}\trm{ ergs}$.\label{fig:Etime}}
\end{figure}

\section{Results and Discussion}
\label{sec:results}

To accurately determine the saturation energy of the $(l = 1, n = 2)$ parent mode, we solve the time-dependent amplitude equations that describe the coupling of the parent mode to the set of high-order g-mode pairs described in \S~\ref{sec:proto}. Figure \ref{fig:Etime} shows the resulting evolution of the parent mode energy for $\dot{E}_1=10^{50}\trm{ ergs s}^{-1}$ and $\dot{E}_1=10^{51}\trm{ ergs s}^{-1}$. 

\begin{figure}
\begin{center}
\epsfig{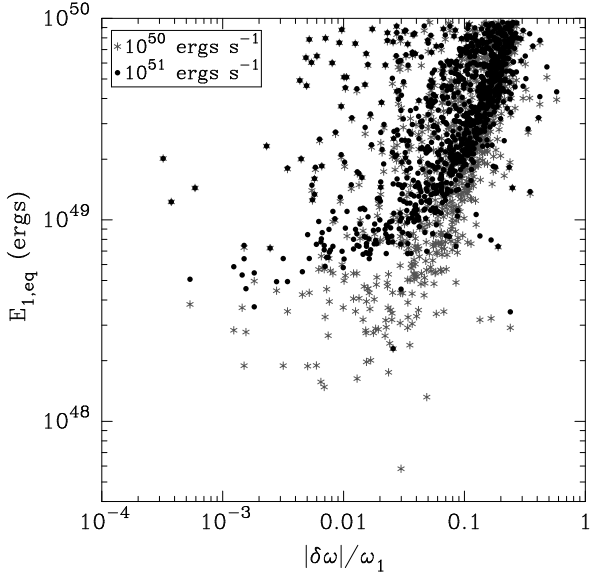}
\end{center}
\caption{Equilibrium energy of the primary g-mode as a function of detuning frequency $|\delta \omega|/\omega_1$ for individual 3-mode coupling between the primary and each g-mode pair with $1\le l \le 10$ and $1/4 \la \omega/\omega_1 \la 3/4$. We show results for $\dot{E}_1=10^{50}\trm{ ergs s}^{-1}$ (\emph{asterisks}) and $\dot{E}_1=10^{51}\trm{ ergs s}^{-1}$ (\emph{circles}). \label{fig:Esat}}
\end{figure}

The parent mode energy saturates at $E_{\rm sat}\approx 1.5\times10^{47}\trm{ ergs}$ for $\dot{E}_1=10^{50}\trm{ ergs s}^{-1}$ and $E_{\rm sat}\approx 2\times10^{48}\trm{ ergs}$ for $\dot{E}_1=10^{51}\trm{ ergs s}^{-1}$.  At these energies, the parent and daughter modes are only mildly nonlinear, with $\delta \rho / \rho \la 0.1$; the nonlinear interactions are therefore well-described by the leading-order nonlinear terms accounted for here. Although the simultaneous interaction between the parent and all daughter pairs is clearly more intricate than individual 3-mode coupling, Figure \ref{fig:Esat} shows that the latter provides a good estimate of the parent's saturation energy. Note that there are dozens of daughter pairs with $E_{1,\rm{ eq}} \sim 10^{48}\trm{ ergs}$ ($5\times10^{48} \trm{ ergs}$) for $\dot{E_1} = 10^{50}\trm{ ergs s}^{-1}$ ($10^{51} \trm{ ergs s}^{-1}$). Parameter values for a few of the low-$E_{1, {\rm eq}}$ daughter pairs are given in Table \ref{tab:tab1}. 

Our calculations imply that most of the energy supplied to the parent core g-mode by accretion onto the PNS will be thermalized as heat via neutrino damping of higher order ($n\ga 10$), parametrically resonant daughter pairs; such short wavelength interactions cannot be resolved in current core collapse simulations.   For a given saturation energy $E_{\rm sat}$, the acoustic power radiated by the PNS is $\sim \gamma_{\rm ac} E_{\rm sat}$, where $\gamma_{\rm ac}$ is the acoustic damping rate of the parent, which Burrows et al. find is $\gamma_{\rm ac} \approx 10\trm{ Hz}$.  For simplicity we assume that $\gamma_{\rm ac}$ is the same at the low saturation energies found here,\footnote{This need not be the case if the coupling between the parent and outgoing sound waves is nonlinear.} in which case the acoustic power radiated is $\sim 10^{48}-2\times10^{49}\trm{ ergs s}^{-1}$ for the values of $E_{\rm sat}$ in Figure \ref{fig:Etime}.  This corresponds to a few percent of the power supplied to the primary core g-mode.  This level of energy injection into the surrounding star is insufficient to drive a core-collapse explosion.  In order for acoustic power to generate a successful $\sim 10^{51}\trm{ erg}$ explosion, we find that $\ga1-3\times10^{52}\trm{ ergs}$ of accretion energy must be supplied to the primary mode. This is $\sim 10-100$ times more energy than is seen in Burrows et al.'s simulations.

The damping rates $\gamma_{2,3}$ and coupling coefficients $\kappa$ are subject to uncertainties that could influence the results presented here.  In particular, determining the exact value of $\gamma_{\rm thin}$ [equation (\ref{eq:gthin})] will require a more careful treatment of neutrino transport near the neutrinosphere.  Uncertainties in the equation of state and the neglect of general relativity and rotation introduce additional uncertainties into $\gamma_{2,3}$ and $\kappa$.   To examine the effect of these uncertainties on the primary's saturation energy $E_{\rm sat}$, we have recalculated the time-dependent amplitude equations with all of our fiducial $\kappa$'s and $\gamma_{\rm thin}$'s multiplied by constant artificial factors.  For $\dot{E}_1 = 10^{51} \trm{ ergs s}^{-1}$, the saturation energy is determined by daughters confined to the core and $E_{\rm sat} \sim \dot{E}_1/\gamma_{\rm thick}$ (the three root solution in \S~\ref{sec:saturation}).   We thus find that even if we multiply/divide our $\kappa$'s and $\gamma_{\rm thin}$'s by factors of $\sim3$, $E_{\rm sat}$ only changes by $20-30\%$.  Since the diffusive approximation is well-satisfied in the inner $\sim 10\trm{ km}$, and  since our fiducial $\gamma_{\rm thick}$'s  [equation (\ref{eq:gthick})] are consistent with a PNS with a Kelvin-Helmholtz cooling time of $t_{\rm KH} \approx 10 \trm{ s}$, we do not believe that $\gamma_{\rm thick}$ is uncertain at the factor of $\sim 3$ level. In particular, decreasing $\gamma_{\rm thick}$ (and thus increasing $E_{\rm sat}$) by more than a factor of $3$ would imply $t_{\rm KH}> 30\trm{ s}$, inconsistent with detailed calculations of PNS cooling (e.g., \citealt{Pons:99}). For $\dot{E}_1 \sim 10^{50}\trm{ ergs s}^{-1}$, daughters confined to the core again determine the saturation energy, but $E_{\rm sat} \sim [18 \kappa^2 Q_2 Q_3]^{-1}$ (the one root solution in \S~\ref{sec:saturation}).  As before, because the core damping is in the diffusive limit, uncertainties in the relevant daughter quality factors are unlikely to be large.   In addition, even a factor of 3 decrease in $\kappa$ only increases $E_{\rm sat}$ to $\sim 10^{48}\trm{ ergs}$. We thus believe that the general conclusion of this {\it Letter} is reasonably robust to uncertainties in PNS microphysics.  In the future, a more detailed mode network will be utilized to further assess this conclusion.
 
Several additional consequences of our results should be noted.  The increase in neutrino flux due to the thermalized pulsational energy is negligible compared to the total flux diffusing out of the PNS. It is therefore unlikely to assist the neutrino mechanism. Our results also imply that g-modes in PNSs cannot be strong sources of gravitational wave radiation. Lastly, future core-collapse simulations must account for sub-grid mode coupling in order to suppress the artificial growth of low-order core oscillations in cases where such growth influences the simulation results. 

\section*{Acknowledgments}
We thank A. Burrows for providing the PNS models and for helpful comments on an earlier draft of this paper, P. Arras for a valuable correspondence on the calculation of coupling coefficients, and T. Thompson, T. van Hoolst, and Y. Wu for helpful suggestions and input. This work was supported by the Theoretical Astrophysics Center at UC Berkeley and by NASA grant NNG06GI68G and the David \& Lucile Packard Foundation.

\label{lastpage}

\end{document}